\documentclass[12pt,a4paper]{article}
\usepackage[T1]{fontenc}
\usepackage[english]{babel}
\usepackage{graphics}
\usepackage{setspace}

\input tcilatex
\begin{document}

\title{Kinetic Monte Carlo Study of Electrochemical Growth in a Hetheroepitaxial
System}
\author{Mar\'{\i}a C. Gim\'enez, Mario G. Del P\'opolo, Ezequiel P. M. Leiva \thanks{Corresponding author. Fax 54-351-4344972;
e-mail: eleiva@fcq.unc.edu.ar}
\\
Unidad de Matem\'atica y F\'{\i}sica, Facultad de Ciencias Qu\'{\i}micas,\\
Universidad Nacional de C\'{o}rdoba, 5000\\
C\'{o}rdoba, Argentina\\
\\
}
\maketitle

\begin{abstract}
\bigskip
Structural and kinetic aspects of 2-D irreversible metal deposition under
potentiostatic conditions are analyzed by means of dynamic Monte Carlo
simulations employing embedded atom potentials for a model system. Three
limiting models, all considering adatom diffusion, were employed to describe
adatom deposition.The first model (A) considers adatom deposition on any free
substrate site on the surface at the same rate. The second model
(B) considers adatom deposition only on substrate sites which exhibit no
neighboring sites occupied by adatoms. The third model (C) allows 
deposition at higher rates on sites presenting neighboring sites occupied 
by adatoms. Under the proper conditions, the
coverage($\theta $) vs time($t$) relationship for the three cases can be
heuristically fitted to the functional form $\theta =1-exp(-\beta t^\alpha
), $where $\alpha $ and $\beta $ are parameters. We suggest that the value
of the parameter $\alpha $ can be employed to distinguish experimentally
between the three cases. While model A trivially delivers $\alpha =1$,
models B and C are characterized by $\alpha <1$ and $\alpha >1$ respectively.

\end{abstract}

\section{Introduction}

The study of the kinetics of formation and growth of two dimensional phases
in electrochemical systems is a classical problem in the field of physical
chemistry. This sort of studies are experimentally performed by applying
various perturbations to the system in order to obtain the response of a
macroscopic observable which allows to infer the distinctive features of the
nucleation and growth process. Over the last two decades, the improvement in
in-situ nanoscopy techniques has also made possible to observe the time
evolution of the surface morphology in electrodeposition experiments as it
is done in the case of deposition experiments under ultra-high vacuum (UHV)
conditions.

 From a theoretical point of view, a large variety of models are generally
used to describe nucleation and crystal growth phenomena \cite
{Barabasi_Libro}. Many of these models apply continuum theories, while
others use descriptions based on microscopic concepts, as does the DDA
(Diffusion, Deposition, Aggregation) model which will be referred to in
section 4. Most of these models have been formulated to describe
experimental situations similar to the ones set up under UHV conditions,
where the adsorption process occurs randomly and at constant flux at any
place on the surface. However, during electrodeposition of a metal, the
entrance of ad-atoms to the system is a thermally-activated process in which
the activation energy for ion reduction is determined mainly by the
reorganization energy of the solvent \cite{Libro_Wolfgang}. This particular
feature has raised a question about the type of site on the electrode
surface where particle deposition occurs. It is possible that in some
systems the reduction reaction \textit{occurs mainly on the edge of
previously formed islands,} whereas in other systems \textit{particles
deposit preferentially onto the clean terraces} and then diffuse in order to
join the growing islands as proposed by Bockris \cite{Bockris_Libro}. A
conclusive answer to this question has not yet been found. This present work
has two main aims. On one hand, we seek to study the most remarkable features
of the electrochemical response and the evolution of surface morphology of
the system for the two limiting cases mentioned above, in comparison with
the trivial case where deposition occurs at the same rate on all free places
of the surface. On the other hand, we shall give simple heuristic criteria
that should allow the identification of the nature of the discharge process
from experimental data.

\section{Approach to the problem}

As a model system, we consider here electrodeposition in the case of a
negligible adsorbate/substrate lattice misfit through the dynamic Monte
Carlo (DMC) method. In order to describe the interaction between particles
in a realistic metallic system, we use semiempirical potentials
corresponding to Ag deposition on Au(100). However, we believe that many
features of the present model will also be valid for other hetheroepitaxy
systems beyond the particular potential employed to describe the interaction
between the particles in the system. Furthermore, the DMC method allows to
study the system evolution in real time. Concerning the type of perturbation
applied to the system, in the present work we shall simulate a
potentiostatic experiment, were the electrode potential is changed from an
initial value where the surface of the electrode is free from adatoms, to a
value where adsorption occurs. Furthermore, we shall consider irreversible
deposition, that is, we neglect adatom dettachment from the surface.

The relationship between the atom deposition rate on a given type of site $x$
and the potential applied is given by the following equation\cite
{Lorenz_Libro}:

\begin{equation}
k_{dep,x}(E)=k_{dep,x}^0a_{Me^{z+}}exp\left( -\frac{\Delta G_{dep,x}^{(0)}}{%
RT}\right) exp\left( -\frac{(1-\alpha _c)zFE}{RT}\right)  \label{k}
\end{equation}
where $k_{dep,x}$ is the atom deposition frequency on site $x$ , $E$ is the
measured-electrode potential vs. a reference electrode, $\Delta
G_{dep,x}^{(0)}$ is the activation energy for ion transfer from the solution
to the crystal at E=0, $\alpha _c$ is the charge transfer coefficient, $%
a_{Me^{z+}}$ is the activity of the metallic ion in the electrolyte, and $%
k_{dep,x}^0$ is the rate constant for the deposition reaction on a site $x$.
According to equation (\ref{k}), a simulation of adatom deposition appears a
priori as a formidable task, since $k_{dep,x}(E)$ should be calculated for
every possible type of site on the surface. This would require a detailed
knowledge of the electron transfer itself. For example, the calculation of $%
k_{dep,x}(E)$ could involve on one hand the local electronic properties of
the site where the discharge of the ion is taking place, as well as the
nature of the rearrangement of the solvent concomitant with the charge
transfer process. However, as we stated in the introduction, in this work we
shall be concerned with some special cases of equation (\ref{k}) and we
shall think of the surface as essentially made of two types of surface
sites: those which are next to at least one adatom and those which are not. Thus,
in principle only two different values of $k_{dep,x}(E)$ may occur. Among
the various possibilities that emerge even after this simplification, the
following three possible mechanisms for atom attachment to the growing phase
were studied, which are considered in Figure 1:

\begin{enumerate}
\item[A]  Particle adsorption occurs on all unoccupied sites on the surface,
which are considered as equivalent, independently of whether or not the site
is surrounded by atoms. In this case, a rate $k$ is assigned to the process
of atom entrance into an unoccupied site, independently of its surroundings.

\item[B]  Particle adsorption occurs only at sites corresponding to the
terraces, i.e. where no adsorbed atoms are around. In this case an
adsorption rate $k$ is assigned to those sites. The adsorption rate is
considered to be $0$ at those sites that have at least one of the nearest
neighboring sites occupied, i.e. on the edge of steps or on kink sites. This
model is based on Bockris' idea that the ion reduces preferentially on the
terraces, since it loses there the least part of its solvation sphere. After
that, the ion diffuses towards the step edges and then towards the kink
sites.

\item[C]  Particles are considered to be discharged preferably at step
edges. With this purpose, a rate $k_1$ is assigned to the entrance into
terrace sites, i.e. where no neighboring atoms are around, and a rate $k_2$
(where $k_2$ \TEXTsymbol{>} $k_1$) is given to the entrance of atoms into
sites that have at least one atom as its nearest neighbor. This situation is
comparable to the Avrami model \cite{Avrami,Libro_Wolfgang} in which the
monolayer grows from the island edges, assuming that atoms enter more easily
there.
\end{enumerate}

In the present model the parameter $k$ was changed between $10^{-3}s^{-1}$
and $10^{+2}s^{-1}$ (at each adsorption site) and the nature of the site $x$
was determined by its atomic enviroment as stated above. \textit{In the present 
model no assumption is made  concerning the evolution of the
surface morphology}. This is a result of the deposition rate $k$ and the
interaction between the particles of the system that affects the diffusion
properties. Thus emphasis is set on making a realistic model of atom
diffusion on the surface, according to a potential that takes into account
the many body effects and which is suitable for the study of metals.

This paper is organized as follows: the model and the simulation technique
used are described in Section 3; the results are given in Section 4, which
is subdivided into two parts: in the first, the system dynamic response is
studied in terms of the evolution of the coverage degree and of the current
(defined as $i=d\theta / dt$)
as a function of time; in the second part, structural aspects are dealt
with, that is, a description of the morphology and quantification of the
formed islands is given. Finally, our conclusions are presented in Section 5.

\section{Lattice Model and Simulation Technique}

\subsection{Dynamic Monte Carlo Method}

Monte Carlo methods are used as computational tools in many areas of
physical chemistry. Although traditionally applied to obtain equilibrium
properties, they can also be used to study dynamic phenomena \cite{MCD}. In
order to do this, the following conditions must be fulfilled: a) in addition
to satisfying the detailed balance criterion, the probabilities of
transition must reflect a ''dynamic hierarchy''; b) time increments between
events must be correctly formulated in terms of the microscopic kinetics of
the system; c) the events must be effectively independent. In the DMC method
every step consists in a random selection of one of the possible processes.
The probability of a process being selected is directly proportional to its
rate. Once the randomly chosen method has been performed, all the possible
processes are calculated and stored again in the corresponding vector and
there is a time increment of $\Delta t=-ln(u)/\sum v_i$ where u is a random
number between 0 and 1, and $\sum v_i$ is the sum of the rates of all the
possible processes. This time increment is due to the assumption that we are
dealing with a Poisson process\cite{MCD}.

Silver adsorption on a defect-free Au(100) surface was studied. The
computational model worked with a square arrangement of $n\times n$
adsorption sites, under periodic boundary conditions. The simulations were
performed with $n=50$ and $n=100$. Unless otherwise stated, the results
presented correspond to the former case. The sites corresponding to the
nearest neighbors on each side were kept in a matrix. The system was
initialized: for the present studies, the initial state corresponded to a
clean Au(100) surface, but the model could be easily extended to surfaces
exhibiting previously adsorbed silver or gold atoms. The rates of every
possible process were stored in a vector to perform the random selection
described above. In this work since we consider the case of irreversible
deposition (no desorption), such processes were the entrance of an atom into
an empty site and the motion of an atom from a site to one of the four sites
corresponding to first nearest neighbors (face 100). For
example a fixed rate may be assigned to the first process, 
independently of the occupation of
the neighboring sites. This would cause an increase in the coverage degree
given by a law of the type $\theta =1-exp(-kt)$, where t denotes the time
elapsed. In the most general case, the entrance rate of a particle depends
on whether or not the site is surrounded by other atoms, as discussed below.

The algorithm developed by Hoshen and Kopelman was used to calculate the
island number and size \cite{Hoshen-Kopelman} and average was taken over 10
simulations for each rate. The interval was divided into 100 bins in order
to calculate the average at each coverage degree.

\subsection{Energy Calculation}

To calculate the activation energies for adatom diffusion the Embedded Atom
Method (EAM) was used \cite{Daw-Baskes}. This method takes into account many
body effects; therefore, it represents better the metallic bonding than a
pair potential does. The total energy of the system is calculated as the sum of
the energies of the individual particles. Each energy is in turn the sum of
an embedding (attractive) energy and a repulsive contribution that arises
from the interaction between nuclei. The EAM contains parameters which were
fitted to reproduce experimental data such as elastic constants, enthalpies
of binary alloys dissolution, lattice constants, and sublimation heats. We
have employed the EAM within a lattice model as described in
references \cite{Ag-Au-thermo,Ag-Au-dyn}.

\subsection{Diffusion Rate Calculations}

Diffusion rates used in the simulation were previously tabulated according
to the different environments that an atom can find on the surface. Various
configurations were generated, consisting of different arrangements of atoms
near the starting and ending sites. For each configuration the path followed
by an atom to jump from a site to the neighboring one was traced and the
energy in each position was calculated, minimizing it with respect to z
(coordinate perpendicular to the surface plane). The activation energy $E_a$
was calculated as the difference between the saddle point and the initial
minimum in the energy curve along the reaction coordinate. The vibrational
frequency $\nu $ of the atom in the starting site was calculated performing
the harmonic approximation near the minimum of the curve. The diffusion rate
was calculated as $v=\nu \ exp(-E_a/kT)$. More details of this calculation
procedure can be found in reference \cite{Ag-Au-dyn}

\section{Results}

\subsection{Evolution of the coverage degree and the current as a function
of time}

\subsubsection{Model A: Adsorption Independent of Surroundings. Dynamic
response.}

Model A is the simplest one and the evolution of the coverage degree
responds to the equation d$\theta $/dt=$k$(1-$\theta $), where k denotes the
adsorption rate per site. In this case $\theta $=$1-exp(-kt)$ and the
current satisfies the law $i=k\ exp(-kt)$. In a plot of the coverage degree
as a function of $kt$, the curves corresponding to the different simulations
overlap, i.e. the evolution of $\theta $ as a function of $kt$ is
independent of the atom entrance rate. However, as we shall later see, the
surface structure differs significantly regarding the island number and
size. Note the difference of the present electrochemical
(potentiostatic)conditions with respect to the UHV ones. In the latter case,
the coverage degree evolves linearly with time, while in the potentiostatic
case the rate is proportional to the available surface resulting in an
exponential law.

\subsubsection{Model B: Preferential Adsorption on Terraces. Dynamic
response.}

For case B, that is, with atom adsorption only on the terraces, Figure 2
shows the evolution of $\theta $ as a function of $kt$ for several $k$
values. Here it can be observed that for higher entrance rates the surface
tends to a lower coverage, i.e. it takes longer to reach a high coverage,
while for lower rates the coverage tends to be closer to one. This is so
because at higher rates more and smaller islands are formed, generating a
larger amount of edge sites where deposition is not allowed. On the other
hand, at low deposition rates the particles have enough time to diffuse
reaching a more compact distribution, i.e. forming fewer and larger islands,
therefore leaving more sites available for adsorption. Further analysis
shows that at relatively short times $\theta $ initially evolves like $%
\theta =1-exp(-\beta t^\alpha )$, with $\alpha $\TEXTsymbol{<}1. This can be
verified by a plot where ln(-ln(1-$\theta $)) is represented as a function
of ln(t) (Figure 3), since straight lines of slope $\alpha $ and y-intercept 
$ln\beta $ are obtained. Table 1 shows $\alpha $ and $\beta $ values for the
various k. For higher rates we found $\alpha $ values considerably lower
than those in case A (equivalent sites) where $\alpha $= 1. For lower rates $%
\alpha $ increases tending to 1. If the system is allowed a longer time to
evolve, a change in the exponent is observed, i.e. the curve becomes much
more horizontal. This effect is much more pronounced at high adsorption
rates. The physical explanation for this is that at a given moment many
small islands have already formed, and consequently few unoccupied sites are
left on the terrace, which makes the entrance of new atoms more difficult.
At this point diffusion starts playing a fundamental role, since the islands
start rearranging in order to form larger ones. This process is slow compared with
the diffusion of free atoms on the terrace. As diffusion occurs, adsorption
sites are generated, allowing a slow entrance of new atoms, which produces a
small current. Finally, at high coverage vacancy islands and simple hollows
are left . There the diffusion rate of vacancies is higher and so is the
generation of adsorption sites, which causes the curve slope to increase at
the final stage.

Figure 4 shows the evolution of $\theta $ and of the current as a function
of time for k = 1 s$^{-1}$. Here a comparison can be made between model A
(adsorption equal at all sites), and model B (adsorption only on the
terraces). In the case of the latter, the coverage degree is found to be
lower and tends to 1 more slowly for the reasons stated before. For the
former case, current decreases exponentially, whereas for adsorption only on
the terraces, it decreases faster at the beginning and more slowly
afterwards, satisfying the law $d\theta /dt=\beta \alpha t^{\alpha
-1}exp(-\beta t^\alpha )$. The points obtained as the result of taking
average over 10 simulations are shown as well as the curves predicted by
this ansatz, generated with the values of $\alpha $ and $\beta $ obtained
from the fit.

\subsubsection{Model C: Preferential adsorption on the Edges. Dynamic
response.}

The third case corresponds to atom adsorption that is preferential on step
edges. Simulations with different adsorption rates on the terraces ($k_1$)
and on the edges ($k_2$) were performed. Table 2 shows the rate constants of all
the systems that have been studied here.

In these simulations the time at which the first particle deposits is taken
as the initial time, since the time elapsed until this first process occurs
may be, in some cases, significantly longer than the time taken by the
simulation, especially when particle adsorption occurs on the edges of
islands at rates much higher than those on the terraces. Figure 5 shows the
evolution of the coverage degree as a function of time for the cases having $%
k_2=1.0s^{-1}$ as entrance rate on the step edges, and $k_1=0.01s^{-1}$, $%
k_1=0.001s^{-1}$ and $k_1=0.0001s^{-1}$ as entrance rates on the terrace. In
these three cases, as well as in all cases where the adsorption rate is
higher on step edges than on the terrace, it can be noticed that the initial
evolution of the coverage degree has a curvature that first points upwards
and then downwards, i.e. it presents a saddle point. This is due to the fact
that the current depends mainly on the amount of edge sites, which initially
increases and afterwards decreases as coalescence occurs.

In these cases, as before, the coverage degree was assumed to satisfy a law $%
\theta =1-exp(-\beta t^\alpha )$ and was represented in a plot as $%
ln(-ln(1-\theta ))$ as a function of $ln(t-t_0)$ in order to obtain the
coefficients $\alpha $ and $\beta $. These are summarized in Table 3. These
values give a qualitative idea of the tendencies, since they correspond to a
single simulation. It can be observed that, in all cases, $\alpha $ is
larger than 1, and it tends to decrease for a given value of $k_1$ as $k_2$
decreases. Figure 6 shows the current as a function of time, along with the
curves $d\theta /dt=\beta \alpha t^{\alpha -1}exp(-\beta t^\alpha )$, with
the parameters obtained from the fit. Three examples are shown. The plot at
the bottom of the page corresponds to $k_1=10^{-4}s^{-1}$ and $%
k_2=10^2s^{-1} $. Here the two rates are very different from each other and
it can be clearly observed that initially the current is 0, then it
increases up to a peak , and after that decreases. The plot in the middle
corresponds to an intermediate case, where the rates are $k_1=10^{-3}s^{-1}$
and $k_2=1.0s^{-1} $ and the curve is qualitatively very similar to that of
the previous case. The plot at the top corresponds to $k_1=10^{-2}s^{-1}$
and $k_2=2\times 10^{-2}s^{-1}$. In this case the two rates are very similar
and a current maximum is not observed in the simulated curves. This case is
more similar qualitatively to the exponential decrease corresponding to
equal adsorption at all sites. $\ $

\subsection{Structural Aspects}

In this section we shall mention first some of the quantities often used to
describe the dynamic behavior of the surface structure in the so-called
atomistic crystal growth models. We shall especially make use of the
concepts used in the DDA model, which considers three basic processes as
responsible for epitaxial growth: Deposition, Diffusion, and Aggregation 
\cite{Family,Bartelt}. Then, we shall study the behavior of these quantities
in each of the three cases considered in this work and their relationship
with the corresponding current transients described in the previous section.

\subsubsection{Use of the Island-Size Distribution Function to Describe
Surface Morphology}

A fundamental quantity in describing kinetically the growth of epitaxial
monolayers and submonolayers is the island-size distribution function $%
N_s(t) $, which is the density per site of islands of size $s$ at time $t$, $%
s$ being the number of atoms in the island. The function $N_s(t)$ is used in
the DDA model to describe the surface structure and it has mainly been
applied to the analysis of experimental results obtained under UHV conditions. In
this case the coverage degree $\theta $ changes linearly with time and then
the function of size distribution is generally expressed as a function of $%
\theta $. In this work we shall follow the same criterion, though we should
keep in mind that in our simulations $\theta $ does not evolve trivially,
because of which many of the relationships found do not strictly correspond
to those of the dynamic scaling.

Defining the total number of islands $N$ and the coverage degree $\theta $
for a given instant by

\begin{equation}
N=\sum_{s\geq 2}N_s\qquad \theta =\sum_{s\geq 1}sN_s
\end{equation}

then the average island size $S(\theta )$ can be written as a function of
the zeroth and first moments ($\sum_{s\geq 2}N_s$, $\sum_{s\geq 2}sN_s$) as

\begin{equation}
S(\theta )=\frac{\sum_{s\geq 2}sN_s}{\sum_{s\geq 2}N_s}=\frac{\theta -N_1}N
\end{equation}

In analyzing the evolution of the surface structure and its relationship to
the evolution of the current and of the coverage degree, we shall
distinguish various dynamic regimes in the behavior of the total island
density $N$ and of the monomer density $N_1$. Then, we shall present our
work in a way similar to the one by Family et al \cite{Family}.

It is important to highlight that the scaling theory is satisfied only under
certain deposition conditions (e.g. relatively low deposition rates), and
naturally, only during the aggregation regime. Thus only some of the
performed simulations can be expected to satisfy completely the observations
predicted by this theory. Such is the case for models A and B when the
coverage degree is approximately at the interval (0.1,0.4) and the
deposition rate is low (low overpotentials), as we shall see in sections
4.2.2 and 4.2.3.

\subsubsection{Model A: Adsorption Independent of Surroundings. Structural
Aspects}

As mentioned in Section 4.1.1, in model A the evolution of the coverage
degree and of the current are given trivially by an exponential function.
However, the observed variety of surface morphology depending on the
deposition conditions indicates that no relationship exists between the
surface structure and the electrochemical response.

Six different entrance rates of atoms were studied: $10^{+2}s^{-1}$, $%
10^{+1}s^{-1}$, $10^0s^{-1}$, $10^{-1}s^{-1}$, $10^{-2}s^{-1}$ and $%
10^{-3}s^{-1}$. Figure 7 shows the evolution of the surface for two
different adsorption rates: the left-hand column corresponds to $%
k=10^{+2}s^{-1}$, and the right-hand column to $k=10^{-2}s^{-1}$. From top
to bottom the coverage degrees are $\theta =0.1$, $\theta =0.3$ and $\theta
=0.6$. As it can be observed, at higher adsorption rates more and smaller
islands are formed. This is true when coverage is low or intermediate, since
at high coverage and rates islands are branched so they coalesce in order to
form one island. At high rates and low coverages a large number of monomers
can be observed. This is due to the fact that, when particles have such a
high entrance rate, they have little time to diffuse, whereas at a low
entrance rate particles have time to join the already existing islands.

Figure 8 shows the total island number, and the monomer number divided into
the total site number for the various entrance rates of atoms. The upper and
the lower part of the figure corresponds to the island density and
to the monomer desity respectively. As it can be observed, at higher deposition
rates both the monomer density and the island density are larger. In all
cases the monomer maximum is reached at lower coverages than the maximum in
the island density. At low deposition rates three zones can clearly be
distinguished during deposition. The first one consists in an initial
increase in the island number, which corresponds to the nucleation process.
Nuclei are formed at the beginning, so we can refer to an instantaneous
nucleation. Then, the number of islands remains essentially constant and
this corresponds to the aggregation regime in which every adatom joins an
existing island whose size increases in this way, but whose number does not.
Finally, the coalescence regime sets in, during which the number of islands
decreases as they begin joining one another to form larger islands.
Eventually, only one island is formed which grows up to form the complete
monolayer. At higher deposition rates, the island density reaches a maximum
at higher coverage degrees, and we can refer to a progressive nucleation,
since the number of islands is increasing over a larger zone of $\theta $
and it is not possible to differentiate a specific aggregation stage.

Figure 9 shows the average size of islands (calculated as the average number
of atoms of every island divided into the total number of sites) as a
function of the coverage degree for the various adsorption rates. In all
cases there is, initially, a linear increase in the island size, with higher
slopes for lower rates. At higher rates, there is a very small increase in
the average island size, since new small islands are continuously forming.
On the other hand, at low rates few islands are formed, whose size increases
considerably as new atoms join them. After that there is an abrupt change in
size, which corresponds to coalescence, and the last stage observed
corresponds to size increase of a single island.

Percolation properties of the system were also analyzed. The percolation
coverage degree $\theta _p$ is defined as the coverage degree at which at
least one island is formed that crosses over the system, at least in one
direction. Figure 10 shows $\theta _p$ as a function of $log(R)$, with $%
R=D/k $, where D is the diffusion rate of the atoms on the surface ($D=436\
s^{-1}$ in this case). This was calculated for a $100\times 100$ surface and
averaged over several simulations. As it can be observed, $\theta _p$
increases as $R$ does, that is, as atom adsorption rate decreases. This is
due to the fact that at lower rates islands are more compact, therefore they
take more time to join one another. On the other hand, at high adsorption
rates islands branch and they are more likely to join together to form a
single island at lower $\theta $'s.

\subsubsection{Model B: Preferential Adsorption on Terraces. Structural
Aspects.}

As it was previously detailed, model B considers that adatom deposition
occurs only at sites with coordination 0. In electrochemical terms, this
means that the reduction reaction occurs only on defect-free terraces. In
this case, we have shown above that the evolution of the coverage degree
(and of the current) is non-trivial and is given by a ''stretched
exponential'' function which emerges from a scenario similar to
that proposed by Palmer et. al. \cite{Palmer,Shlesinger}. The 
exponent $\alpha$ depends on the applied
overpotential,  and  in all cases is less than one. Simulations
performed under conditions analogous to those of Figure 7 yielded images that
are qualitatively very similar to those of model A presented there. However,
it was found that for model B the time at which each degree of coverage
occurs is longer, that is to say, the filling in of the surface is slower.
Comparative figures are given in Table 4. It can be noticed that this
effect is stronger at higher coverages and at high deposition rates. As
explained before this is due to the lesser availability of adsorption sites.

In Figure 11 we can observe the evolution of the total density of islands $N$
and the density of monomers $N_1$ as a function of $\theta $. A comparison
between this plot and the one of the previous model in Figure 8 shows that
at high rates the island and monomer density are larger, while at low rates
the curves practically agree. This is due to the fact that at high
adsorption rates atoms have little time to diffuse and reorder, and in turn
those that adsorb do it in sites not surrounded by atoms. All this causes
the formation of a larger amount of islands, whereas at low rates, in spite
of adsorption only at neighbor- free sites atoms have sufficient time to
diffuse and join the existing islands and the net growth occurs in a way
similar to that of the previous model. In addition to the larger island
density we can observe a shift at high coverage (between 0.4 and 0.6), which
indicates that there are more islands in that zone, that is to say, the
islands take longer to coalesce and do it at higher $\theta $s.

Figure 12 shows the average size of islands as a function of the coverage
degree. If we compare this plot with the one of Model A we can observe that
the abrupt change in size shifts to higher $\theta $s, especially for the
case of an atom entrance rate of $10^{-3}s^{-1}$, which is the slowest rate.
This is due to island compactness, which causes islands to take longer in
joining and forming a single one.

\subsubsection{Model C: Preferential Adsorption on the Edges. Structural
Aspects.}

This model considers that atoms on island edges adsorb at a rate higher than
that of atoms which adsorb on the terrace. Adsorption rates of $%
10^{-2}s^{-1} $, $10^{-3}s^{-1}$ and $10^{-4}s^{-1}$ on the terrace ($k_1$),
and up to $10^{+2}s^{-1}$ on the edges($k_2$) were studied. For very large
rate differences a single island is formed in our simulation system, which
grows from the edges. As adsorption rates become similar the formation of
more islands is observed. Figure 13 shows the state of the surface at $%
\theta =0.1 $ for different values of $k_1$ and $k_2$, entrance rate on the
terrace and on the edges respectively. The left-hand column corresponds to $%
k_2=10^{+2}s^{-1}$. Under these conditions, only for $k_1=10^{-2}s^{-1}$ two
islands were formed, while at lower values of $k_1$ a single island 
appears. On the other hand, the results on the right column show that for
similar values of $k_1$ and $k_2$ several islands are formed. In all cases,
however, the island shape is quite compact and dendritic shapes are not
observed, as they were found at high rates in the other models.

\section{Conclusions}

In this paper we considered several aspects of 2-D metal deposition by means
of dynamic Monte Carlo simulations As a model system we considered $Ag$
deposition on Au(100) under various deposition conditions using interatomic
potentials adequate for metals and a realistic diffusion simulation model.

The reduction of $Ag^{+}$ ions was modeled in three different ways: in the
first and simplest case, deposition of atoms was assumed to occur on any
unoccupied site of the lattice independently of the site surroundings. Here
the evolution of the coverage as a function of time was given in a trivial
way by an exponential function, while the structure of the surface depended
on the deposition rate. In the second case, particle adsorption was allowed
only on defect-free terraces, that is, only on those sites characterized by
adsorbate-free surroundings. It was observed that the evolution of the
coverage rate is given by a function $\theta =1-exp(-\beta t^\alpha )$,
where the exponent $\alpha $ depends on the deposition rate, that is, on the
overpotential applied, though in all the cases considered it turned out to
be smaller than one. At high deposition rates a change in this exponent was
also observed, which indicates different types of evolution of the surface
structure. Thus a relationship exists between the coverage evolution - and
therefore the current evolution - and the change in the surface structure.
In the third deposition model, the ion reduction was considered to be
possible on edge sites as well as on sites located on clean terraces. A
higher deposition rate was always used on the edges. In this case, the
coverage degree seems to follow again a $\theta =1-exp(-\beta t^\alpha )$
law, but with exponent $\alpha $ always larger than one. This kind of
behavior can be compared with the predictions of the Avrami model \cite
{Libro_Wolfgang}, where the exponent $\alpha $ is two or three, depending on
whether the nucleation is progressive or instantaneous. As regards the
evolution of surface morphology, there was a remarkable reduction in the
number of islands, and a single island was formed in our simulation system
when there was a very large difference between the rates over the two types
of sites. The exponent values reported must be considered to give a
qualitative tendency, in the sense that we did not study systematically the
finite-size effects and they correspond to a single simulation run. 
These features may be important in determining the exact
values of the exponents.

\section{Acknowledgements}

We thank CONICET, SeCyt-UNC, Agencia C\'{o}rdoba Ciencia, Program BID
1201/OC-AR PICT $N^o$ 06-04505 and Fundaci\'{o}n Antorchas (M.G.D.) for
financial support. We also thank to E. Albano and R. Moneti for
Hoshen-Kopelman subroutine. Part of the present calculations were performed
on a Digital workstation donated by the Alexander von Humboldt Stiftung,
Germany. Language assistance by Karina Plasencia is gratefully acknowledged.


\newpage


\section{Tables}

Table 1

Fitted $\alpha $ and $\beta $ values from simulations with various
adsorption rates $k$ onto an unoccupied site . These parameters were
obtained by fitting of the law $\theta =1-exp(-\beta t^\alpha )$ from
simulations with model B described in the text.
\newline

\begin{tabular}{|l|l|l|}
\hline
$k(rate)$ & $\alpha $ & $\beta $ \\ \hline
$10^2$ & 0.722 & 12.72 \\ \hline
$10^1$ & 0.748 & 2.914 \\ \hline
$10^0$ & 0.799 & 0.588 \\ \hline
$10^{-1}$ & 0.868 & 0.095 \\ \hline
$10^{-2}$ & 0.909 & 0.012 \\ \hline
$10^{-3}$ & 0.951 & 0.001 \\ \hline
\end{tabular}
\vspace{1cm} \\ 

Table 2

Parameters employed in the simulation studies with model C. $k_1$ and $k_2$
denote the adsorption rates on sites without adsorbed neighbors( 'terrace
site ') and with adsorbed neighbors( 'border site ') respectively. Crosses
indicate the sets of values of rate constants considered .
\newline
 
\begin{tabular}{|l|l|l|l|l|l|l|l|}
\hline
$k1(terrace)\backslash k2(border)$ & $10^2$ & $10^1$ & $10^0$ & $10^{-1}$ & $%
2\times 10^{-2}$ & $10^{-2}$ & $2\times 10^{-3}$ \\ \hline
$10^{-2}$ & x & x & x & x & x &  &  \\ \hline
$10^{-3}$ & x & x & x & x &  & x & x \\ \hline
$10^{-4}$ & x & x & x & x &  & x &  \\ \hline
\end{tabular}
\vspace{1cm} \\ 

Table 3

Fitted $\alpha $ and $\beta $ values for various simulation conditions with
model C. They are obtained by fitting of the law $\theta =1-exp(-\beta
t^\alpha )$ .
\newline

\begin{tabular}{|l|l|l|l|}
\hline
$k1(terrace)$ & $k2(border)$ & $\alpha $ & $\beta $ \\ \hline
$10^{-2}$ & $10^2$ & 2.3 & 63 \\ \hline
$10^{-2}$ & $10^1$ & 2.4 & 1.3 \\ \hline
$10^{-2}$ & $10^0$ & 1.9 & 0.03 \\ \hline
$10^{-2}$ & $10^{-1}$ & 1.5 & 0.005 \\ \hline
$10^{-2}$ & $2\times 10^{-2}$ & 1.2 & 0.006 \\ \hline
$10^{-3}$ & $10^2$ & 2.1 & 36. \\ \hline
$10^{-3}$ & $10^1$ & 2.0 & 0.51 \\ \hline
$10^{-3}$ & $10^0$ & 2.0 & 0.009 \\ \hline
$10^{-3}$ & $10^{-1}$ & 1.7 & $7.1\times 10^{-4}$ \\ \hline
$10^{-3}$ & $10^{-2}$ & 1.3 & $4.3\times 10^{-4}$ \\ \hline
$10^{-3}$ & $2\times 10^{-3}$ & 1.1 & $7.4\times 10^{-4}$ \\ \hline
$10^{-4}$ & $10^2$ & 2.0 & 27. \\ \hline
$10^{-4}$ & $10^1$ & 1.9 & 0.24 \\ \hline
$10^{-4}$ & $10^0$ & 2.0 & 0.002 \\ \hline
$10^{-4}$ & $10^{-1}$ & 2.1 & $2.9\times 10^{-5}$ \\ \hline
$10^{-4}$ & $10^{-2}$ & 1.7 & $4.9\times 10^{-6}$ \\ \hline
\end{tabular}
\\ 

\newpage

Table 4
Time $t_\theta $ at which a given degree of coverage $\theta $ occurs in
models A and B described in the text. $k$ denotes the rate assigned to the
process of atom entrance onto an unoccupied site.
\newline

\begin{tabular}{|l|l|l|l|l|}
\hline
$\theta $ & $\ 
\begin{array}{c}
t_\theta (k=10^2s^{-1}) \\ 
\func{mod}\text{el A}
\end{array}
$ & $
\begin{array}{c}
t_\theta (k=10^2s^{-1}) \\ 
\func{mod}\text{el B}
\end{array}
$ & $
\begin{array}{c}
t_\theta (k=10^{-2}s^{-1}) \\ 
\func{mod}\text{el A}
\end{array}
$ & $
\begin{array}{c}
t_\theta (k=10^{-2}s^{-1}) \\ 
\func{mod}\text{el B}
\end{array}
$ \\ \hline
0.1 & 0.00094 & 0.0013 & 11.3 & 11.6 \\ \hline
0.5 & 0.007 & 0.020 & 65.5 & 88.2 \\ \hline
0.7 & 0.012 & 0.083 & 114. & 177. \\ \hline
0.9 & 0.024 & 194. & 221. & 430. \\ \hline
\end{tabular}

\section{Figure Captions}

Figure 1: Scheme of the mechanisms for atom attachment to the growing phase
2-D studied in the present work:

\begin{itemize}
\item[A)] Particle adsorption occurs on all unoccupied sites on the surface at the
same rate $k.$
  
\item[B)] Particle adsorption occurs only at sites where no adsorbed atoms are
around ( 'terrace sites ') at the rate $k$. The adsorption rate on sites
that have one or more neighboring adatoms is $0$.

\item[C)] Particles are considered to be discharged preferably at sites with
neighboring adatoms ( 'edges '). A rate $k_1$ is assigned to the entrance
onto terrace sites and a rate $k_2$ ( $k_2$ \TEXTsymbol{>} $k_1$) is given
to atoms entrance into sites that have at least one neighboring adatom.
\end{itemize}

Figure 2: Evolution of the coverage degree $\theta $ as a function of $k*t$
corresponding to the adsorption model B described in the text ('terrace '
adsorption). The adsorption rate $k$ is given in the figure in units of $%
s^{-1}$.

\vspace{1cm}

Figure 3: Plot of $ln(-ln(1-\theta ))$ as a function of $ln(t)$ at various
deposition rates $k$ for Model B described in the text ('terrace '
adsorption). The adsorption rate $k$ is given in the figure in units of $%
s^{-1}$. Slopes (exponents) associated to the linear relation found during
the first stage of deposition are detailed in table 1.  

\vspace{1cm}

Figure 4: Evolution of the coverage degree $\theta $ and of the current for
models A and B for  a deposition rate of $1\ s^{-1}$ . 

Black solid line: evolution of $\theta $ in model A. Gray dashed line:
evolution of $\theta $ in model B.  

Circles: current evolution in model A. The gray solid line represents the
theoretical fitting according to the equation $d\theta /dt=\beta \alpha
t^{\alpha -1}exp(-\beta t^\alpha )$

Diamonds: current evolution in model B. The black dashed line represents the
theoretical fitting according to the equation $d\theta /dt=\beta \alpha
t^{\alpha -1}exp(-\beta t^\alpha )$

The exponents $\alpha $ and $\beta $ are given in Table 1.

\vspace{1cm}

Figure 5: Evolution of $\theta $ as a function of time for the deposition
model C. In these simulations,  a  deposition rate $k_2$ = $1\ s^{-1}$ is
considered on the edges and different deposition rates $k_1$ are assumed at
the terraces. The $k_1$ values are  given in the figure in units of $s^{-1}.$
\vspace{1cm}

Figure 6: Current ($d\theta /dt$) as a function of time for deposition model
C. The full line are simulation results, and the dashed lines show the
curves predicted by the function $d\theta /dt=\beta \alpha t^{\alpha
-1}exp(-\beta t^\alpha ).$ Each plot employs the parameters obtained through
minimum square fits of the simulation data. The adsorption rates employed
for the simulations were the following: Top box:  $k_1=10^{-2}s^{-1}$ ;  $%
k_2=2\times 10^{-2}s^{-1}$; middle box:  $k_1=10^{-3}s^{-1}$ ;  $k_2=1s^{-1}$%
; box at the bottom: $k_1=10^{-4}s^{-1}$ ; $k_2=10^2s^{-1}$.

\vspace{1cm}

Figure 7: Frames showing surface morphology in model A at three different
values of $\theta $. The deposition rates were $k=$ $10^2s^{-1}$ for the
left-hand column and $k=10^{-2}s^{-1}$ for the right-hand column.The coverage degrees
are, from top to bottom, $\theta =0.1$, $\theta =0.3$ and $\theta =0.6$.

The associated times are $t=9.4\times 10^{-4}s$, $t=3.5\times 10^{-3}s$ and $%
t=9.2\times 10^{-3}s$ from top to bottom for the left column and   $t=11.31s$%
, $t=34.22s$ and $t=87.26s$ for the right one.

\vspace{1cm}

Figure 8: Plot of island density (top) and monomer density (bottom) as a
function of the coverage degree for different adsorption rates according to
model A.  Rates used in each case are indicated in the body of the figure in
units of $s^{-1}.$

\vspace{1cm}

Figure 9: Average island size $<s>$ (defined as the average number of atoms
per island divided by the total number of sites $N_T$ ) as a function of the
coverage degree $\theta $, for different adsorption rates according to model
A. The rates used in the simulations are included in the figure in units of $%
s^{-1}$. $N_T=2500.$

\vspace{1cm}

Figure 10: Coverage degree at which percolation occurs, $\theta _p$ as a
function of $log(R)$, where $R=\frac Dk$, $D$ being the diffusion rate of an
adatom on the clean surface and $k$ the particle entrance rate. Simulations
were performed on a $100\times 100$ square lattice.

\vspace{1cm}

Figure 11: Plots of  island density (top) and  monomer density (bottom) as a
function of coverage degree for different adsorption rates according to
model B. Rates used in each case are indicated in the body of the figure in
units of $s^{-1}.$

\vspace{1cm}

Figure 12: Average island size $<s>$ as a function of the coverage degree $%
\theta $, for different adsorption rates according to model B. The rates
used in the simulations are included in the figure in units of $s^{-1}.$

\vspace{1cm}

Figure 13: Frames showing the surface structure in simulations with model C
at a coverage degree $\theta =0.1$ for different adsorption rates on
terraces and edges. Frame 1: $k_1=10^{-2}s^{-1}$;  $k_2=10^2s^{-1}$; Frame
2: $k_1=10^{-2}s^{-1}$; $k_2=2\times 10^{-2}s^{-1}$; Frame 3: $%
k_1=10^{-3}s^{-1}$ ; $k_2=10^2s^{-1}$; Frame 4: $k_1=10^{-3}s^{-1}$;  $%
k_2=10^{-2}s^{-1}$; Frame 5: $k_1=10^{-4}s^{-1}$;  $k_2=10^2s^{-1}$; Frame
6: $k_1=10^{-4}s^{-1}$ ;  $k_2=10^{-2}s^{-1}$.

\newpage

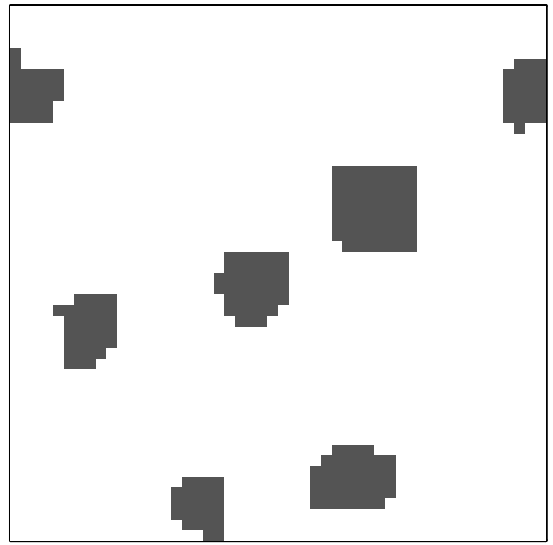
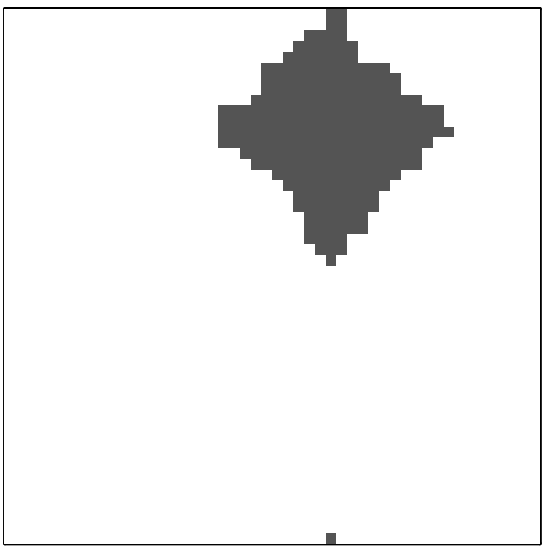
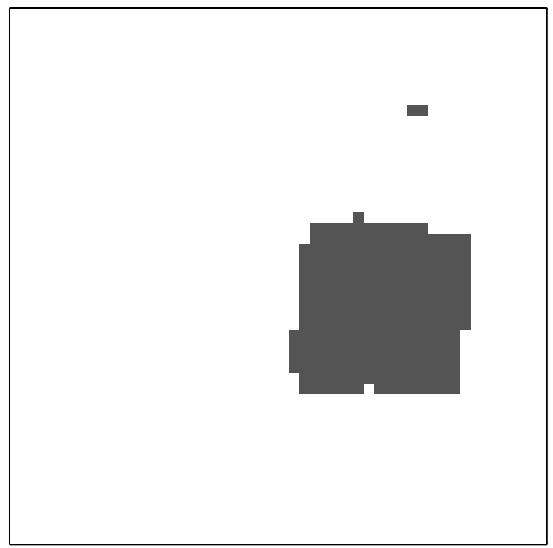
\end{document}